\documentclass[english]{webofc}
\usepackage[varg]{txfonts}   % Web of Conferences font
\usepackage{hyperref}        % hyperref package
\usepackage{epstopdf}        % hyperref package

\begin{document}
\title{The Southern Hemisphere Hunt for Dark Matter at the Stawell Underground Physics Laboratory}
%
% subtitle is optionnal
%
%%%\subtitle{Do you have a subtitle?\\ If so, write it here}

\author{Phillip Urquijo\inst{1}\fnsep\thanks{\email{purquijo@unimelb.edu.au}} 
%\and
%        Second author\inst{2}\fnsep\thanks{\email{Mail address for second
%             author if necessary}} \and
%        Third author\inst{3}\fnsep\thanks{\email{Mail address for last
%             author if necessary}}
        % etc.
}

\institute{School of Physics,
The University of Melbourne,
Parkville, Victoria 3010, Australia }

\abstract{%
I report on the Stawell Underground Physics Laboratory (SUPL), a new facility to be built in 2016, located 1 km below the surface in western Victoria, Australia. I will discuss the status of the proposed SABRE experiment, which will be comprised of a pair of high purity 50-60 kg NaI crystal detectors with active veto shielding to be located in labs in the Northern and Southern Hemispheres respectively. I also discuss projects beyond SABRE, including directional dark matter detectors, which will be used to determine the origin of any true dark matter signals.
\par}
\maketitle
\section{Introduction}
\label{intro}
An understanding of the nature of dark matter is one of the most important challenges faced by particle- and astro- physics today. About 30\% of the total energy density in our Universe is believed to be in the form of relic massive particles interacting only feebly with ordinary matter. One experimentally allowed way for these particles to interact is via the weak interaction, similar to a neutrino. Direct detection of so-called WIMPs (Weakly Interacting Massive Particles) is possible by measuring the nuclear recoil energy following an elastic WIMP scattering off nuclei.

Over the past decade, several dark matter direct detection experiments have found intriguing anomalies, which may be due to WIMP interactions through the combined motion of the Earth in the solar system and in the Galaxy. The most striking of these  results was from the DAMA/LIBRA experiment~\cite{Bernabei:2008yh}, which claimed 9$\sigma$ evidence for annual modulation due to a dark matter hypothesis~\cite{Bernabei:2008yi}.

The Stawell Underground Physics Laboratory (SUPL) will be key in confirming or refuting the claims of DAMA/LIBRA, as host to the southern hemisphere experiment of the SABRE twin experiment. SUPL is the first deep underground laboratory in Australia, and the southern hemisphere. It will be host to dark matter experiments, and to other deep underground science research. In this proceedings I review the phenomena of dark matter annular modulation, the plans for SUPL, SABRE, and directional experiment R\&D.

\section{Dark Matter Observables}\label{sec-1}
Dark matter searches are at the interface of particle-, nuclear- and astro-physics. In the case of WIMP dark matter, the differential nuclear recoil event rate with respect to energy due to dark matter interactions can be written as 
\begin{eqnarray}
\frac{dR}{dE} (E, t) =  \frac{\rho_0}{m_\chi \cdot m_A} \cdot \int v\cdot f({\bf v},t) \cdot \frac{d\sigma}{dE}(E,v) d^3v
\end{eqnarray} 
where $m_\chi$ is the dark matter mass and $d\sigma/dE(E,v)$ its differential cross section~\cite{Undagoitia:2015gya}. The WIMP cross section, $\sigma$,  and $m_\chi$ are the two observables of the experiment. The dark matter velocity $v$ is defined in the rest frame of the detector and $m_A$ is the target nucleus mass. The astrophysical parameters are the local dark matter density, $\rho_0$, and $f({\bf v},t)$, which accounts for the WIMP velocity distribution in the detector reference frame. 

Experimentally, both the energy dependence and the annual modulation of the dark matter signature are analysed. The WIMP-nucleus interaction rate is expected to modulate yearly due to the Earth's rotation around the Sun. The speed of dark matter particles in the Milky Way halo relative to the Earth is largest around June 2nd and smallest in December. Consequently, the number of particles able to produce nuclear recoils above the detectors' energy threshold is also largest in June. As the amplitude of the variation is expected to be small, the temporal variation of the differential event rate can be written as
\begin{eqnarray}
\frac{dR}{dE} (E, t) \approx S_0(E) + S_m(E)\cdot \cos\left( \frac{2\pi (t-t_0)}{T} \right)
\end{eqnarray} 
where $t_0$ is the phase expected at about 150 days and $T$ is the expected period of 1 year. The time-averaged event rate is denoted by $S_0$, and the modulation amplitude by $S_m$.

\section{Previous measurements}
Owing to the very low expected event rate, about $10^{-2}$ events/kg/day for standard assumptions on the WIMP mass (100 GeV), experiments aiming at a direct dark matter search have very stringent background requirements \cite{Undagoitia:2015gya}. Nuclear recoils induced by dark matter interactions are in an energy region heavily affected by environmental radioactive, seasonal, and cosmic background. The only way to adequately reduce the flux of cosmic ray-induced particles is to build the experiment in  deep underground.

\begin{figure}[htbp] 
  \centering
  \includegraphics[width=0.9\linewidth]{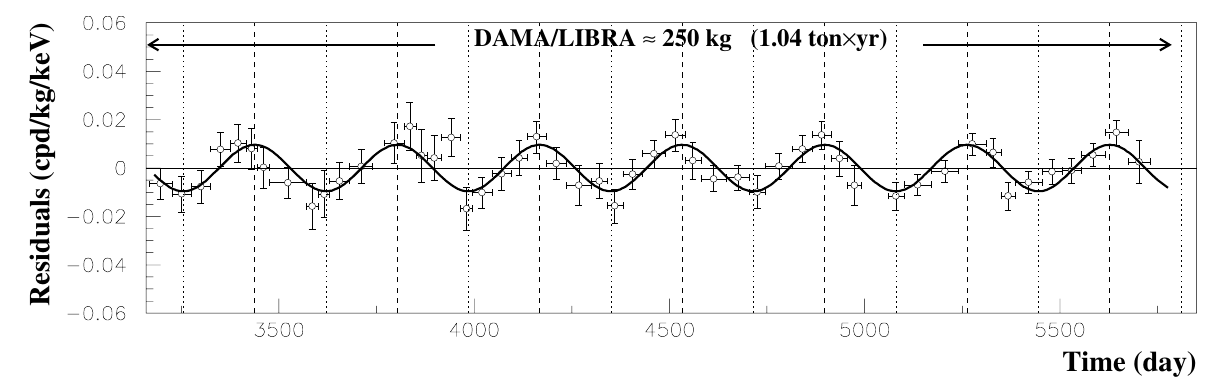}
  \caption{The annual modulation of the low energy event rate in the DAMA/LIBRA experiment. Figure taken from Ref. \cite{Bernabei:2008yi}.}
  \label{fig:dama}
\end{figure}

For over a decade, the DAMA experiment (DAMA-NaI and DAMA/LIBRA), situated at Gran Sasso National Laboratory (LNGS), has been observing a low energy (2-6 keV$_{ee}$ or keV electron equivalent) rate annular modulation in an array of high purity NaI(Tl) crystal detectors, shown in Fig. \ref{fig:dama}.  This modulation occurs with a phase and period consistent with predicted dark matter interactions. Dark matter of similar mass and nuclear cross-section was also recently suggested by the CoGeNT\cite{Aalseth:2010vx}, CDMS-Si, \cite{Agnese:2013rvf} and CRESST II \cite{Angloher:2011uu} experiments.  However, the dark matter interpretation of the DAMA/LIBRA modulation remains in question. Its tension with other experiments such as XENON \cite{Aprile:2012nq}, LUX \cite{Akerib:2013tjd} and KIMS \cite{Kim:2014toa} suggests the need for further investigation of the DAMA modulation.  So far there are no firm explanations for the modulation other than a dark matter hypothesis. Although limits from large noble gas experiments, such as XENON and LUX, are very powerful in constraining dark matter, the WIMP interaction processes they are sensitive can differ to DAMA/LIBRA, and therefore cannot rule it out in a model independent way.

%No experiment, however, has been able to test the DAMA experiment such that the interpretation is completely independent of the dark matter model chosen. 
%It could be that the signal observed by DAMA/LIBRA is related to atmospheric muons, the rate of which is annually modulated due to temperature variations in the stratosphere, or to combinations of muons and modulated neutrinos caused by the varying Sun-Earth distance. Some varying rates of background neutrons have also been considered. Some of these proposals have been refuted but the signal remains and its interpretations remains a puzzle.

\begin{figure}[htbp] 
  \centering
  \includegraphics[width=0.6\linewidth]{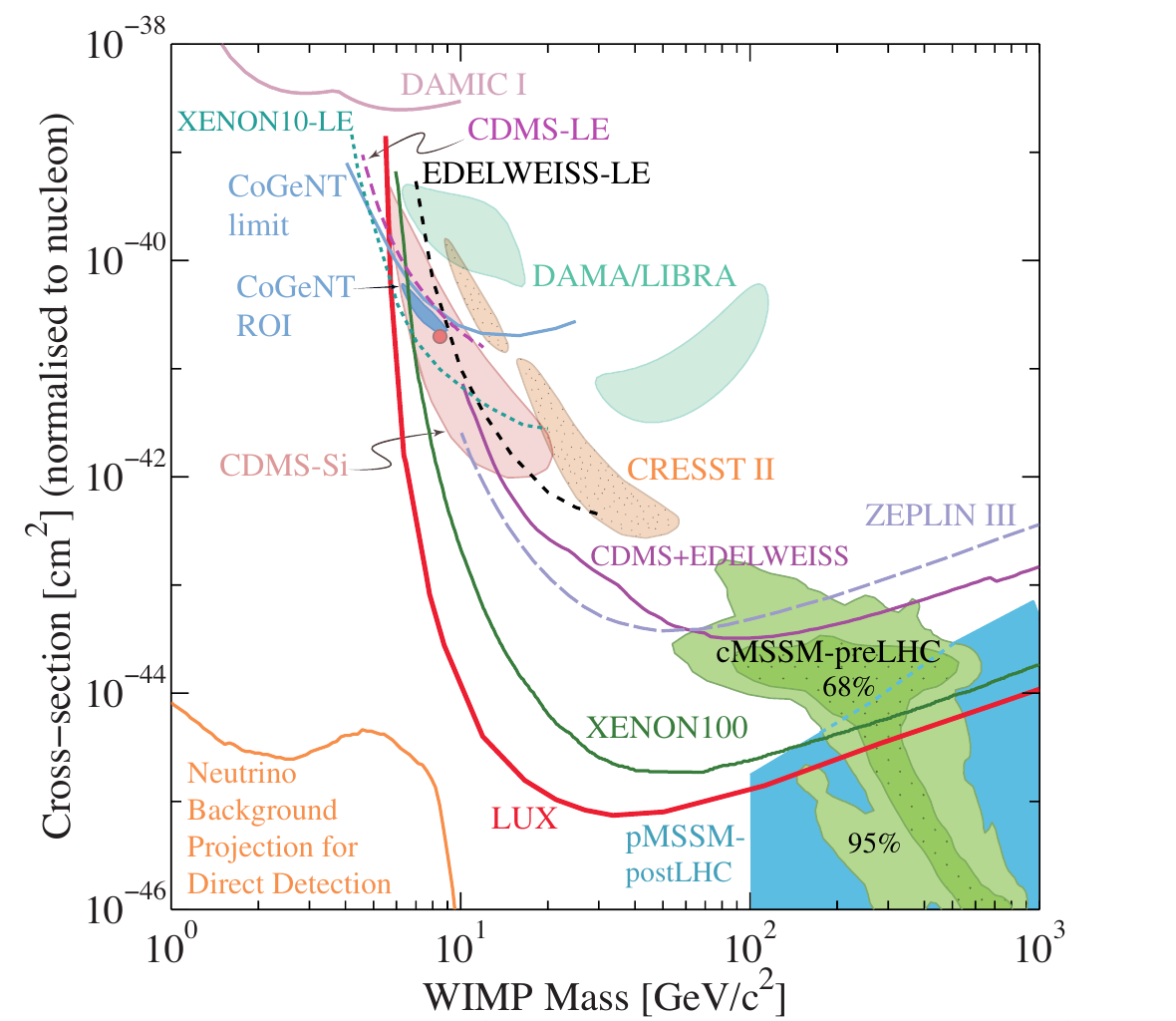}
  \caption{WIMP cross sections (normalised to a single nucleon) for spin-independent coupling versus mass, taken from Ref. \cite{Agashe:2014kda}.}
  \label{fig:limits}
\end{figure}

In general, the differential WIMP-nucleus cross section, $d\sigma/dE$ can be written as the sum of a spin-independent (SI) contribution and a spin-dependent (SD) one,
\begin{eqnarray}
\frac{d\sigma}{dE} = \frac{m_A}{2\mu^2_A v^2} \cdot \left( \sigma_0^{\rm SI} \cdot F_{\rm SI}^2 (E) + \sigma_0^{\rm SD} \cdot F_{\rm SD}^2 (E)\right)
\end{eqnarray}
where $F_i$ are the wimp-nucleon form factors, $\sigma_{0i}$ are the cross sections at zero momentum transfer, $m_A$ is the target mass, and $\mu_A$ is the WIMP-nucleus reduced mass.  The form factors can be determined in nuclear shell model calculations. The spin independent cross sections are typically simple functions of the nucleon number, while the spin dependent cross sections have most recently been determined using chiral effective-field theory and depend on isotope.  
A summary of the current status of dark matter searches is shown in Fig. \ref{fig:limits} for spin-independent limits..

There are other scattering processes that may be allowed by the current experimental data. Various models allow dark matter scattering off electrons. For example, sub-GeV dark matter particles could produce detectable ionisation signals.  If new forms of couplings are introduced to mediate dark matter - electron interactions, further models become viable. Assuming an axial-vector coupling, dark matter -lepton interactions dominate at tree level and cannot be probed by dark matter-baryon scattering.  Models such as kinematic-mixed mirror dark matter \cite{Foot:2014xwa} also predicts interactions with atomic electrons.

Although alternative models may still be able to reconcile DAMA/LIBRA signals with other experimental results, a model-independent test using ultra high purity NaI(Tl) crystal detectors is necessary to confirm or refute the DAMA/LIBRA dark matter claim. The SABRE collaboration proposes to test the DAMA signal in a dual site approach at LNGS (the site of DAMA/LIBRA), and in Stawell. The experiment will consist of highly pure NaI(Tl) crystals in an active liquid scintillator veto to tag and reduce the potassium background. Two identical detectors will be made, for the two sites. This approach allows for substantial mitigation of season modulation effects, particularly seasonal temperature variations. 

\section{Low background physics}
The allowed cross section of a WIMP interaction is extremely low. In order to identify unambiguous signs of dark matter particles, ultra-low background experimental conditions are required. Background from $\gamma$-rays, and neutrons must be mitigated to great extent. The dominant radiation from gamma-decays originates from uranium and thorium chains, as well as common isotopes,  $^{40}$K, $^{60}$Co and $^{137}$Cs present in the surrounding materials. The $^{232}$Th and $^{238}$U chains have a series of alpha and beta decays accompanied by the emission of several gamma rays with energies from tens of keV up to 2.6 MeV. Many of the possible interactions of gamma rays with matter can deposit energy in the target medium in the range of a few keV, which is the sensitive region of detectors like DAMA/LIBRA. To stop $\gamma$ activity from outside of the experimental setup, large water tanks are often employed, although not ubiquitously. They provide homogenous shielding, and satisfy most background requirements. This demands large experimental halls, many times greater than the dimensions of the active detector region. 

Neutrons can interact with the nuclei in the target via elastic scattering, mimicking the nuclear recoil signature of a WIMP interaction. The most challenging case is elastic scattering, as opposed to in-elastic scattering, which is usually accompanied by gamma emission that can be tagged by inducing scintillation in multiple crystals. Most cosmogenic neutrons are produced due to spallation reactions of muons on nuclei in the experimental setup or in the surrounding rock.  While spallation neutrons can be several GeV in energy, subsequent interaction with the detector material reduces the energies to MeV ranges, which can produce nuclear recoils in the dark matter search regime. The most powerful way of moderating the flux of this background is to place the detectors as deep as possible. Radiogenic neutrons must be controlled with good material selection, and a good screening facility.

\section{Stawell Underground Physics Laboratory}\label{sec-1}
An Australian dark matter program has recently been established, to provide unique insight into this puzzle, and to make important contributions to the field of low background physics.  The proposed laboratory will be located at the Stawell Gold Mine, Stawell, Victoria, Australia.  The mine is 250 km west of Melbourne. The Stawell Gold Mine is a decline mine with a helical access road descending the main ore body with many large caverns dug into the ore body along the road. It is a dry basalt rock mine with a flat overburden, and reaches a depth of 1.6 km. The mine is still active.

\begin{figure}[htbp] 
  \centering
  \includegraphics[width=0.6\linewidth]{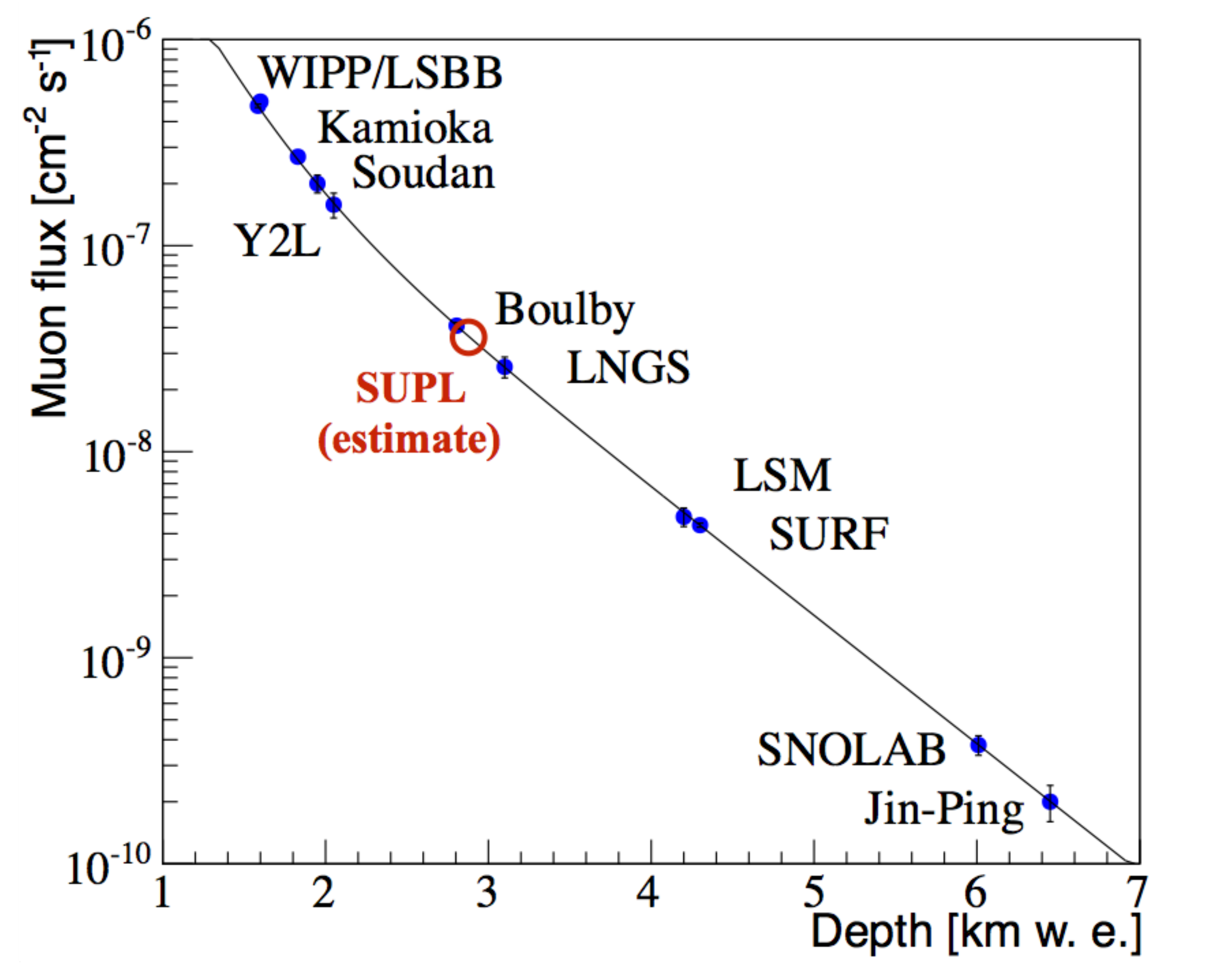}
  \caption{Muon flux as a function of depth in km of water equivalent for various underground laboratories. The SUPL depth and muon flux is approximate pending precision results. The parameterisation (solid line) is taken from Ref \cite{Mei:2005gm}.}
  \label{fig:supldepth}
\end{figure}

\begin{figure}[htbp] 
  \centering
  \includegraphics[width=0.6\linewidth]{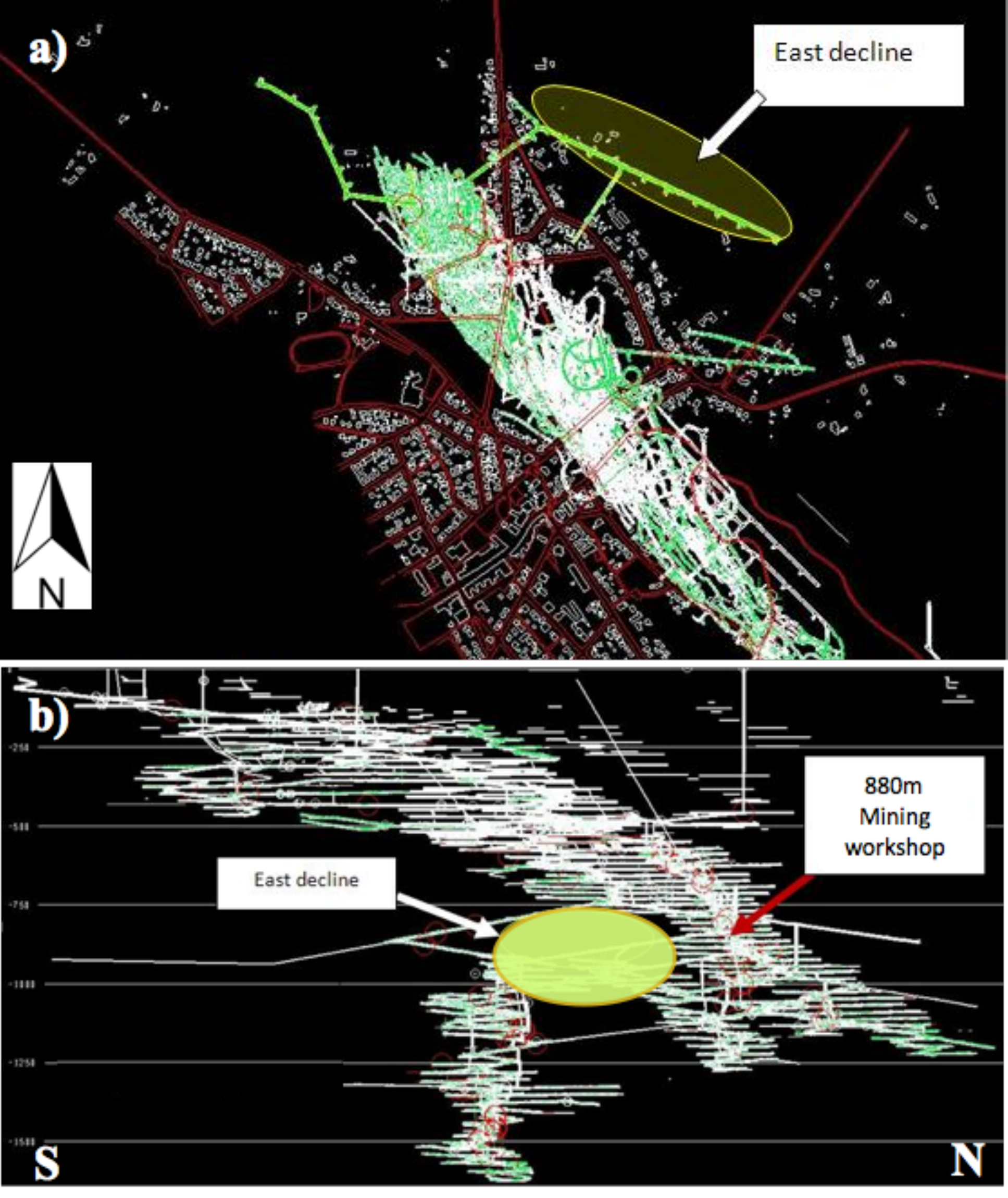}
  \caption{The proposed location of SUPL in the Stawell Gold Mine: a) Topological view of Stawell and the mine tunnels underneath, b) side view of the tunnels on the North-South projection. The laboratory will be located near the east decline.}
  \label{fig:lablocation}
\end{figure}

The location and the nature of the mine make it ideal for dark matter experiments. We have identified a site at a depth of 1025 m to host the laboratory. Given the high density of basalt, 2.86 g/cm$^3$, the site has approximately 2900 m water equivalent of overburden, similar to the average depth of the Gran Sasso National Laboratory in Italy (Fig. \ref{fig:supldepth}).  It is situated near the east decline, where there are no mining tunnels overhead, shown in Fig. \ref{fig:lablocation}.
The laboratory is planned to contain a dark matter experiment and an integrated underground laboratory facility for hosting other research. 

A series of preliminary characterisation measurements have been carried out at a depth of 880 m to estimate the conditions of the laboratory. These are being repeated near the nominal laboratory site. The neutron flux was found to be approximately $7 \times 10^{-6}$ n s$^{-1}$ cm$^{-2}$ for fast neutrons (0.5 MeV and above) and  $2 \times 10^{-6}$ n s$^{-1}$ cm$^{-2}$ for slow neutrons (< 0.5 MeV), however a dedicated program of precise neutron energy spectrum measurements is underway to inform shielding design. The upper limit on the $\gamma$ ray flux for energies below 3 MeV is 2.5 $\gamma$ s$^{-1}$ cm$^{-2}$. The surrounding rock contained approximately 0.6 ppm $^{238}$U and 1.6 ppm $^{232}$Th. The $^{40}$K concentration was found to be as low as 20 Bq/kg.
These conditions have been found to be  within acceptable tolerances as defined by LNGS, $i.e.$ similar to Gran Sasso. The muon, slow-neutron, fast-neutron and gamma ray fluxes are currently being measured with high precision to inform detector shielding design.  A new muon detector setup has been built in Melbourne to ensure low noise, and maximum muon purity. The first results from the upgraded muon detector are due in early 2016. The radon levels are acceptable when ventilation is provided to the experiment area, and will be further controlled within each experiment.

\begin{figure*}[htbp] 
  \centering
  \includegraphics[width=0.8\linewidth]{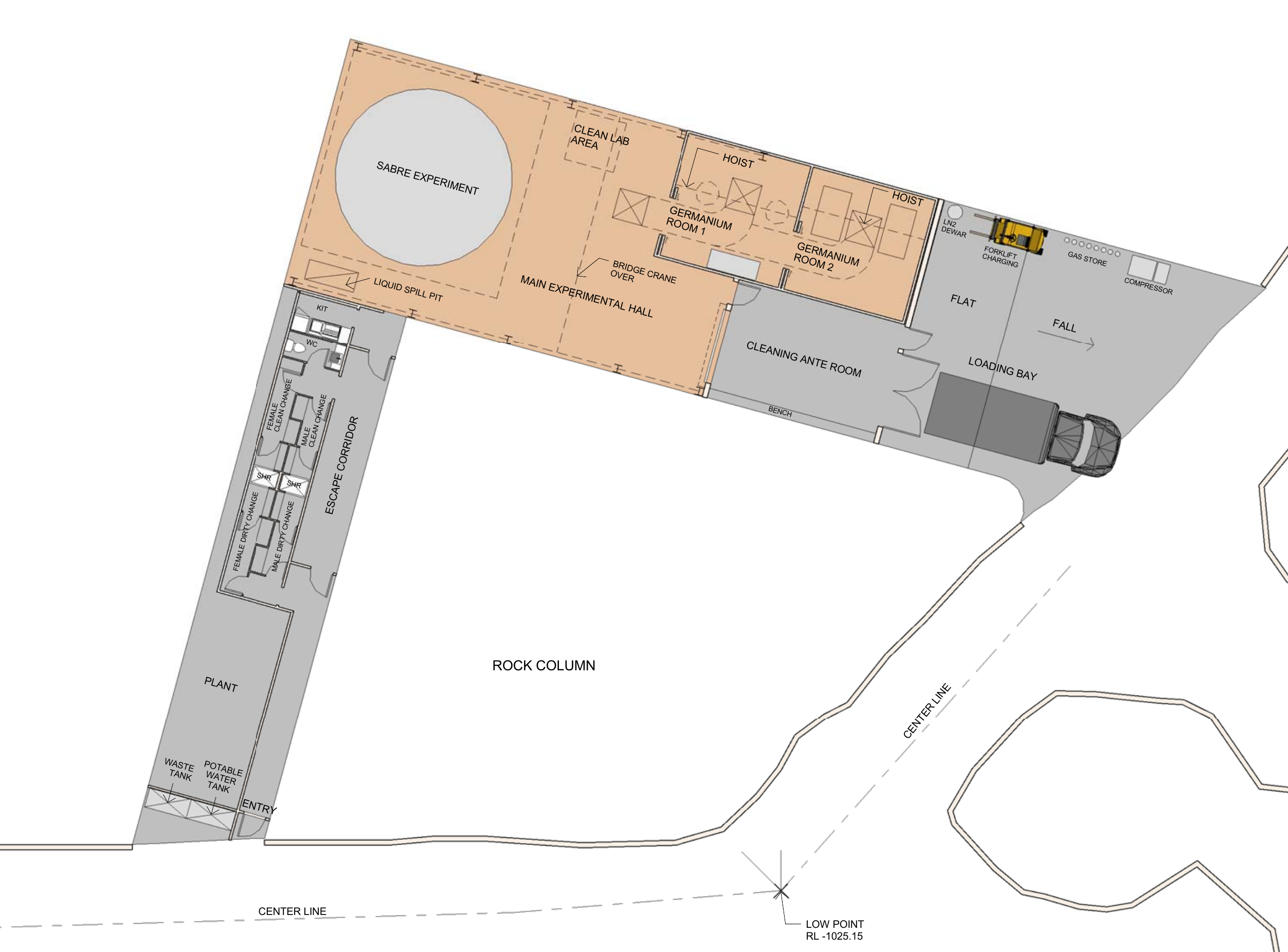}
  \caption{Conceptual design for the SUPL laboratory, located 1025 m below the surface.}
  \label{fig:supldesign}
\end{figure*}

Preliminary plans for the design of SUPL have been drafted, and the engineering company responsible for leading the construction has been tendered. The laboratory space will be built during 2016, for delivery in early 2017. The design is to have a main cavity of 35-40 m (length) $\times$ 10 m (width) $\times$ 10 m (height), with a loading area of 15 m (length), and a class 1000 clean area of 25 m (length), shown in Fig. \ref{fig:supldesign}. It will house a clean ante room and washing area, a separate high grade clean room, and two low background screening facilities. The design of the laboratory will allow for a possible expansion space in the future. 

SUPL is planned to be a multi-use underground laboratory. Its flagship will be the SABRE dark matter detector, but other uses will include nuclear screening, astrobiology, and as a host for research and development of new underground experiments in dark matter and neutrino physics.

\section{SABRE}\label{sec-1}
The SABRE (Sodium iodide with Active Background REjection) experiment seeks to conduct a direct test of DAMA/LIBRA through a number of innovations. The key features are the development of ultra-high purity NaI(Tl) crystals and the rejection of the residual $^{40}$K background (among others) using an active veto. High-purity crystals have been developed at Princeton. The active veto has been designed with a liquid scintillator vessel. 
 
 \subsection{Experiment plan}
The final SABRE setup will be two identical experiments with 50-60 kg of NaI crystals. A prototype, or commissioning detector will be conducted with a small 1-2 kg crystal.  Each crystal will be enclosed in a self-contained module consisting of the crystal itself and two low-radioactivity, high quantum efficiency photomultiplier tubes. The phototubes are optically coupled to opposite sides of the crystal. These components are then sealed in a light and air tight low radioactivity metal container.  The crystal detectors are placed inside of a cylindrical liquid scintillator, 1.5 m diameter and 1.5 length, detector that is shielded from environment background with passive shielding. The passive shielding is still be designed for the respective locations at Gran Sasso and Stawell. The liquid scintillator detector is able to reject internal radioactivity as well as external $\gamma$ ray backgrounds.  The scintillator vessel will contain approximately 2 tonnes of linear alkylbenzene (LAB). Approximately twenty 20.2 cm Hamamatsu R5912 PMTs will be placed at the ends of the cylindrical vessel.  

The most powerful, and novel aspect to the SABRE experiment is the dual location approach. By placing one detector in the southern hemisphere at Stawell ($37^\circ 03'$ S) and one in the northern hemisphere at Gran Sasso ($42^\circ28'$ N), the seasonal thermal impact on background to dark matter annual modulations can be constrained. Dark matter candidates from the galactic halo wind will however provide consistent, in phase, signatures in the two locations.  Each detector will use the same technology and detection approach, thereby reducing model dependence.

\subsection{Active veto concept}
\begin{figure}[htbp] 
  \centering
  \includegraphics[width=0.6\linewidth]{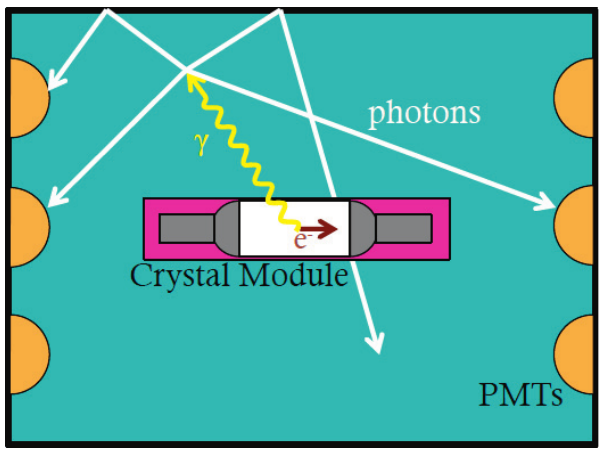}
  \caption{Illustration of the veto concept, demonstrating the detection of a $^{40}$K decay X-ray/Auger electron and corresponding 1.4 MeV $\gamma$ ray. The active veto observes the $\gamma$ ray in the liquid scintillator material and the events are rejected.}
  \label{fig:vetoscheme}
\end{figure}

Sodium Iodide crystals, such as those used in DAMA/LIBRA contain traces of $^{40}$K (a 13 ppb average is reported). When $^{40}$K decays to $^{40}$Ar, it produces a 1.46 MeV $\gamma$-ray at the same time as a possible 3 keV X-ray/Auger electron. %, shown in Fig. \ref{fig:kdecay}. 
If the $\gamma$ ray escapes the detector, the 3 keV deposition will become a source of background in the primary energy region of interest. Simultaneous, separate detection of these two signals, however, can be taken as a signature of $^{40}$K decay and can be used to reject such background events. DAMA/LIBRA suppressed some of the $^{40}$K background by rejecting multi-hit events in which more than one crystal sees an event. However, this approach was inefficient, as evidenced by the $\sim3$ keV$_{ee}$ peak in their energy spectrum.
%\begin{figure}[htbp] 
%  \centering
%  \includegraphics[width=0.6\linewidth]{potassiumdecay.pdf}
%  \caption{The decay scheme of $^{40}$K.}
%  \label{fig:kdecay}
%\end{figure}
In the SABRE design, a dedicated detector surrounds the crystal modules, which can be used to detect radiation leaving the dark-matter sensitive volume. The outer detector is filled with liquid scintillator, LAB, and provides an active veto for radioactive background events such as $^{40}$K decay. An illustration of the principle is shown in Fig. \ref{fig:vetoscheme}.

\subsection{Crystal production}
The most important part of the experiment is the reduction of $^{40}$K contamination in the scintillator crystals.  The Princeton group partnered with Sigma-Aldrich to produce NaI powder with lower background than previously achieved. The primary aim has been to reduce $^{40}$K levels below the 13 ppb seen by DAMA/LIBRA. There is also concern for background from U, Th, and Rb, which all contribute to the flat background in the energy spectrum.  Sigma-Aldrich has made a high purity ``Astro-grade" powder with a reported K level of 3.5 ppb. This powder has been used to grow test crystals, and is the nominal choice for the final full scale experiment.

DAMA/LIBRA has a single interaction rate of $\sim$1 cpd/kg/keV$_{ee}$, with reported crystal radioactivity levels of 13 ppb nat K (average) and $\sim$1-10 ppt for U and Th. The SABRE crystals are currently being grown from the ``Astro-grade" powder. The scintillation yield and optical properties of the crystal are also very important: they are directly related to the detected signal amplitude, and thus to the achievable energy threshold of the experiment. Large crystal masses are also necessary for the experiment to achieve a high dark matter sensitivity. Due to light attenuation effects, longer crystals decrease the light yield, and crystals with large cross-section are preferred. We expect to grow NaI(Tl) crystals around $10.2$ cm in diameter and $25.4$ cm long with a total mass of 8-10 kg each. 
%The Kyropoulos process is the preferred crystal growth method for large crystals, and will be used for SABRE. It has been shown that from a concentration of about 10 ppb K in the crystal, an event rate of $^{40}$K of around 1 cpd/kg is expected. Therefore, given the recent improvements to the purity in the powder, 
It is expected that the SABRE crystals will have much less than 1 count per day per kg.

\begin{figure}[htbp] 
  \centering
  \includegraphics[width=0.6\linewidth]{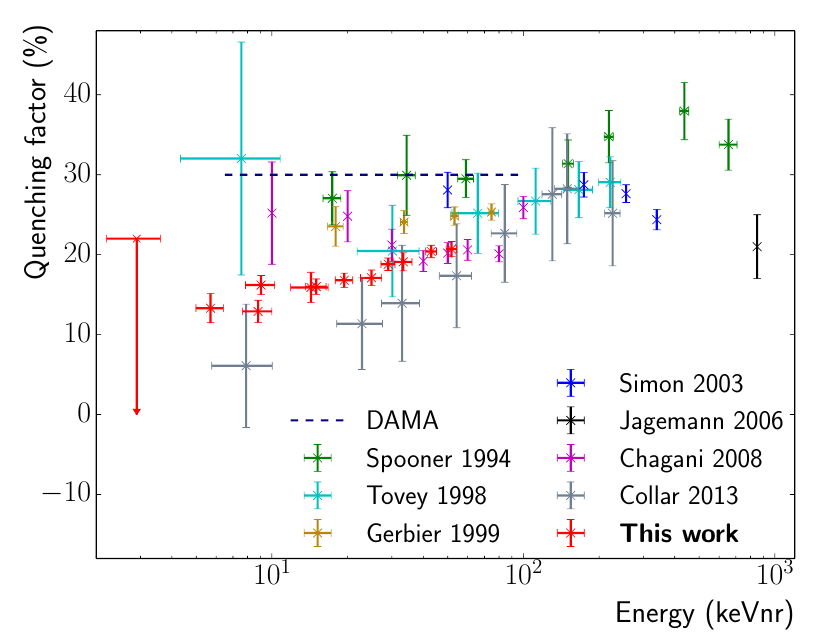}
  \caption{Measurements of the Na-recoil quenching factors as a function of recoil energy. The figure is taken from Ref. \cite{Xu:2015wha}, and the most recent measurement shown with red markers.}
  \label{fig:quench}
\end{figure}

The dark matter interpretation of the DAMA/LIBRA (or SABRE) modulation signal depends on the scintillation efficiency of NaI(Tl) for sodium and Iodine recoils relative to that produced by WIMP scattering interactions. Nuclear recoils in NaI(Tl) tend to produce less scintillation light than electron recoils with the same energy deposition.  The relative ratio is usually referred to as the quenching factor.  DAMA reports a quenching factor of 0.3 for sodium and 0.09 for iodine recoils, by exposing the NaI(Tl) crystals to a $^{252}$Cf source. A more recent analysis by members of  SABRE have found contradictory results that change in WIMP interpretation \cite{Xu:2015wha}. They use a source of low-energy, pulsed neutrons and combine time-of-flight and pulse-shape-discrimination techniques to perform a very precise measurement of the Na-recoil quenching factors. The results are shown in Fig. \ref{fig:quench}. The implications for DAMA/LIBRA are that the observed modulation signal occurs in the energy window of 13 keV to 32 keV instead of the 7 keV to 20 keV as previously quoted, shifting the WIMP mass interpretation to higher mass values.

\subsection{Measurement expectations}
\begin{figure}[htbp] 
  \centering
  \includegraphics[width=0.6\linewidth]{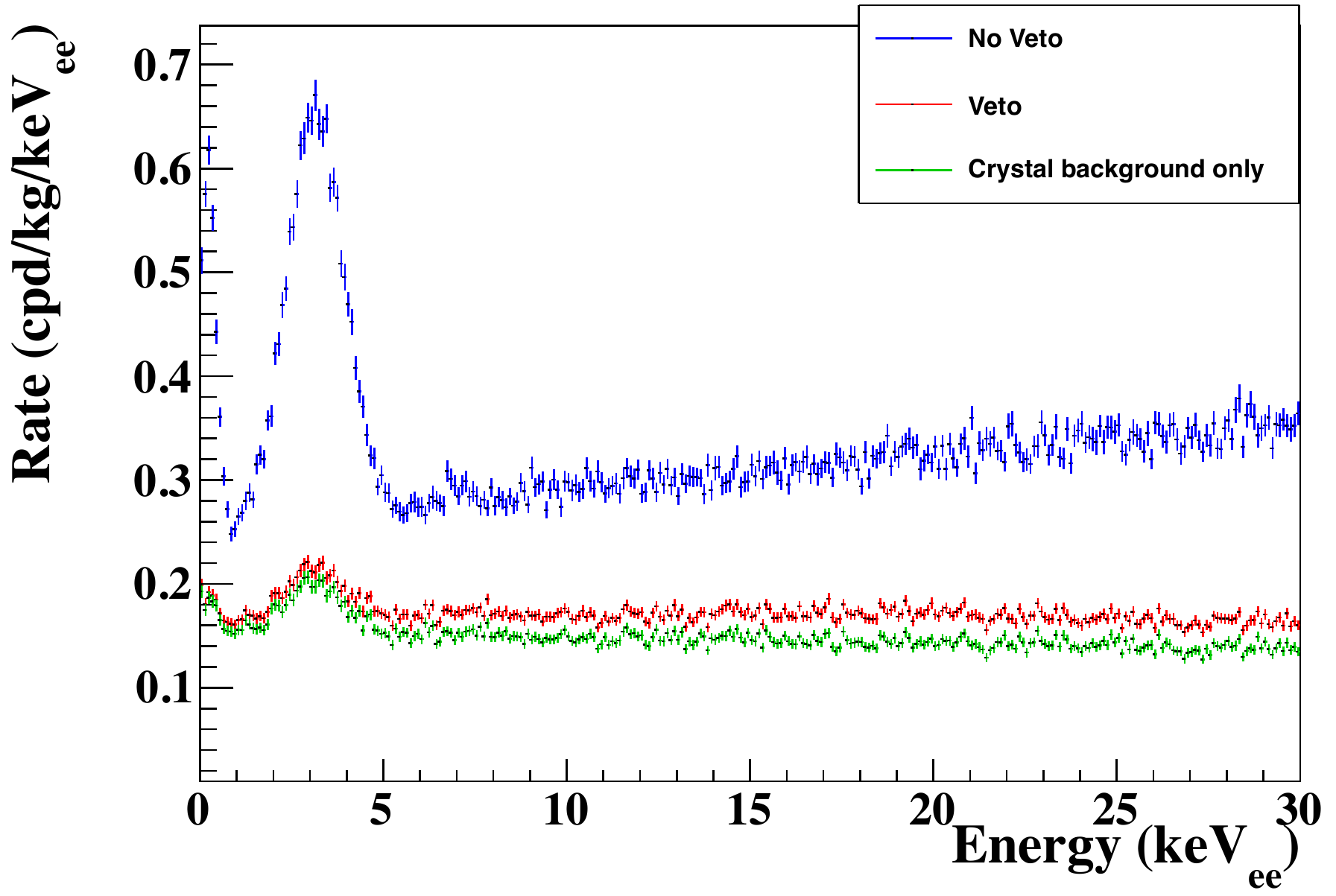}
  \caption{Estimated background in SABRE based on Geant4 Monte-Carlo simulation.}
  \label{fig:sensitivity}
\end{figure}

An estimate of the background for a dark matter measurement has been performed using Monte Carlo simulation (MC) of a 2 kg crystal and assuming DAMA-level background in the crystal \cite{Shields:2015wka}. This simulation was used to demonstrate the veto detector's rejection efficiency.  It was found that of all $^{40}$K events depositing 1-4 keV energy in the crystal, nearly 90\% are rejected by the veto or coincidence with other crystals, shown in Fig. \ref{fig:sensitivity}. It is expected that the SABRE background level would be about two times less than the DAMA/LIBRA flat singles rate. With the conservative predicted background rate of 0.4 cpd/kg/keV$_{ee}$, described in Ref. \cite{Shields:2015wka}, MC studies show that it would take SABRE 3 years to refute or confirm the DAMA/LIBA modulation signal with $>3\sigma$ confidence level. The low radioactivity background achieved by SABRE should also allow it to probe new frontiers in WIMP dark matter searches.

\section{Directional Experiment Prospects}
A positive signature from SABRE must be accompanied with a search program to further elucidate the nature and source of the dark matter signature. Directional dark matter detectors offer an excellent way to achieve this goal. The principle of direct detection is to measure both the energy and the trajectory, of a recoiling nucleus following a WIMP scattering process. The double differential spectrum is dependent on the WIMP velocity distribution, and hence the Milky Way halo model. The WIMP event distribution is expected to present an excess in the direction of motion of the Solar system, which happens to be roughly in the direction of the constellation Cygnus. The WIMP induced recoil distribution presents a dipole feature in galactic coordinates, whereas the background distribution is expected to be isotropic in the galactic rest frame. 
\begin{figure}[htbp] 
  \centering
  \includegraphics[width=0.6\linewidth]{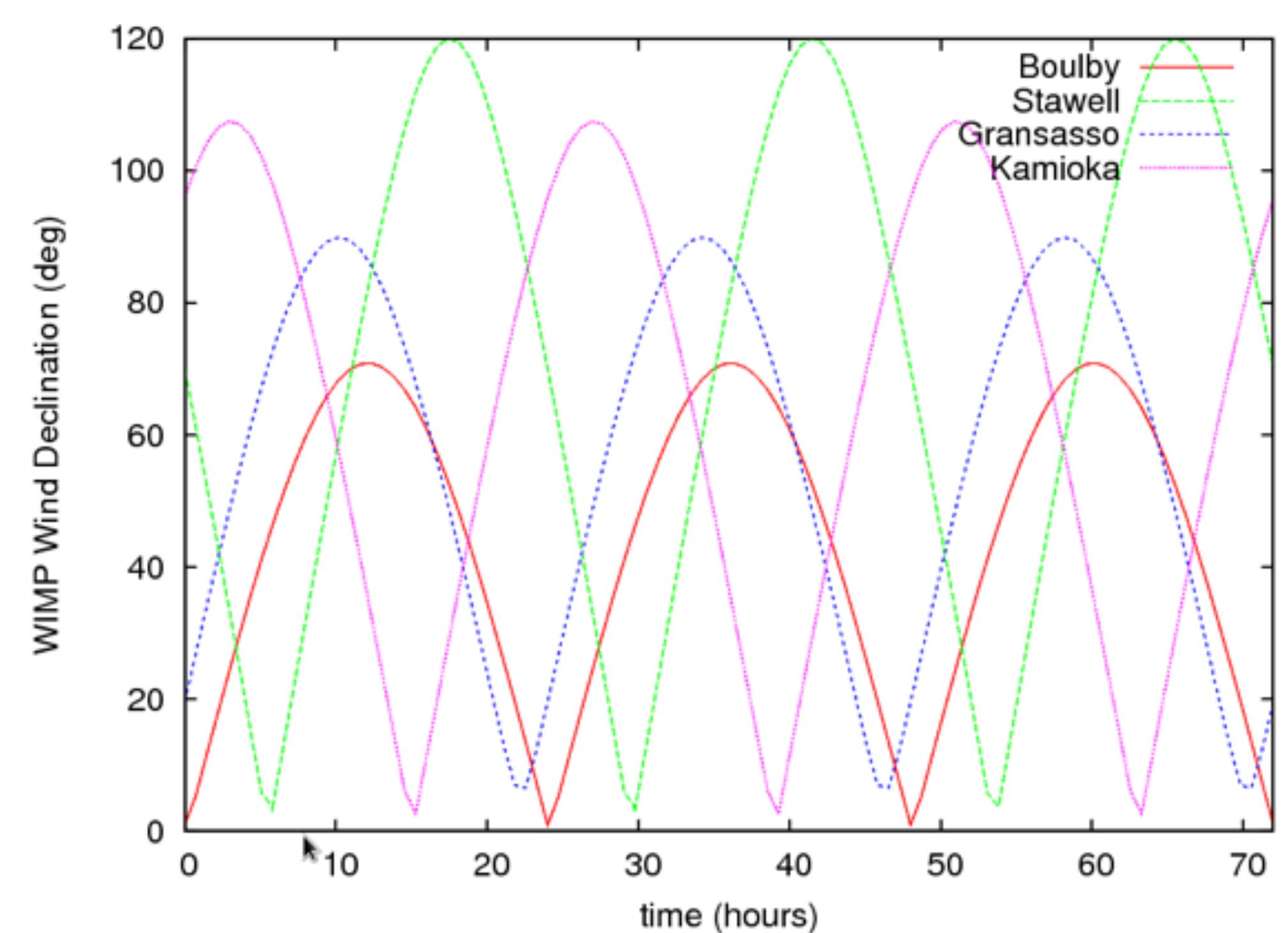}
  \caption{Declination angle to the WIMP wind (Cygnus).}
  \label{fig:declination}
\end{figure}
Stawell sits in a geographically advantageous location for directional dark matter studies, as the angle of inclination to the WIMP wind oscillates with the greatest amplitude of any underground laboratory (Fig. \ref{fig:declination}). This means the experimental signature will be the most pronounced. 

The experimental signature of direct detection techniques is the recoil track produced in the dark matter interaction. This provides information on the ionisation density as a function of position, on the range and eventually on the direction of the recoiling nuclei. The most promising strategy for directional searches is the use of low pressure gas, leaving tracks of 1-2mm in length. The most successful direct detection experiment is DRIFT-II, operated at Boulby \cite{Battat:2014van}. Its volume is 0.8 m$^3$, and is filled with a low pressure mixture of CS$_2$:CF$_4$:O$_2$ gas.  It contains 32.2 g of fluorine in the active volume for WIMP interactions. Research is ongoing with the CYGNUS collaboration to adapt this technology to very large scales, up to 64 m$^3$, and is a candidate project for SUPL in the near future.

\section{Summary}
SUPL will be a major new facility for Australia and for global underground science projects, to be constructed in 2016/2017. The flag-ship experiment, SABRE, aims to test the claims of the DAMA/LIBRA collaboration in a model independent way using sites in the north and south hemispheres. SUPL also offers opportunities for directional dark matter studies, which will be the only way to elucidate the nature of dark matter once found.

\end{document}